\documentstyle[11pt,newpasp,twoside,epsf,psfig]{article} 
\def\edcomment#1{\iffalse\marginpar{\raggedright\sl#1\/}\else\relax\fi} 
\reversemarginpar

\begin{document} 

\title{Spectral Energy Distributions of Blazars: Facts and Speculations} 
\author{Laura Maraschi \& Fabrizio Tavecchio} 
\affil{Osservatorio Astronomico di Brera, Via Brera 28 20121 Milano I}     
 
\begin{abstract} 
We discuss the present knowledge about the Spectral Energy Distributions
(SEDs) of blazars within a unified approach emphasizing overall
similarities.  The properties of the average SEDs of different samples of
blazars suggest that more powerful sources contain on average less
energetic particles.  Detailed studies of TeV emitting blazars show that
the energy of the particles emitting the bulk of the power increases
during flares.  A framework for a general theoretical understanding is
proposed.  We present recent results on the SEDs of a group of blazars
with emission lines, allowing to estimate both the luminosity in the jet
and the luminosity of the accretion disk. Implications for the origin of
the power carried by relativistic jets are considered.
\end{abstract}

\section{Introduction} 

It is now generally accepted tha the blazar "phenomenon" (highly
polarised and rapidly variable radio/optical continuum) is due to a
relativistic jet pointing close to the line of sight.  This concept,
introduced by Blandford and Rees in 1978, represented a fundamental break
through.  It took however a long time to gather sufficient data to apply
it consistently and quantitatively to individual objects and to object
classes.  Taxonomy has confused the issues, since observational
definitions were based (as usual in astronomy) on analogies with known
objects rather than on well defined intrinsic properties. Moreover
variability could alter the "classification" of the same object. 
After decades of studies
to distinguish between objects with or without emission lines (flat
spectrum quasars vs. BL Lacs) or between radio bright and X-ray bright BL
Lacs, we think that presently the most productive approach is to assume
that all blazars contain relativistic jets and ask in what way these jets
differ in different objects and eventually why.

Therefore here and in the following we will assume that Quasars with Flat
Radio Spectrum (FSQs, which include OVVs and HPQs) and BL Lac objects are
essentially "similar" objects in the sense that the nature of the central
engine is similar apart from some basic scales.  The obvious parameters
are the central black hole mass, angular momentum and the accretion rate
(Blandford 1990). We don't know yet what governs the phenomenology but we
assume that it is some combination of these three parameters.  The goal
is to understand the role of these fundamental parameters starting from a
physical comprehension of the phenomenology. We will therefore take 
a particular "point of view" and not attempt a review of all the work
carried out in this field.

\section{The average SEDs of different samples of blazars.}
 
It was noted early on that the SEDs of blazars exhibited remarkable
systematic properties (Landau et al. 1986, Sambruna et al. 1996).  The
subsequent discovery by the Compton Gamma Ray Observatory of gamma-ray
emission from blazars (a summary can be found in Mukherjee et al. 1997)
was a major step forward, showing that in many cases the bulk of the
luminosity was emitted in this band and questioning the importance of
previous studies of the SEDs at lower frequencies.

Although a large fraction of the sky was surveyed with the EGRET
instrument on board CGRO, its sensitivity was just sufficient to detect
the brightest or flaring sources. Therefore the results did not allow to
define a complete sample.

A simple approach was taken by Fossati et al. (1998) in order to explore
systematically the properties of the blazar SEDs up to gamma-ray
energies. The procedure was to construct "average" SEDs of known complete
samples of blazars (FSRQs from the 2 Jy radiosample, BL Lacs from the 1 Jy
radiosample and BL Lacs from the Slew Survey X-ray sample) using
published fluxes at 6 wavelengths in the radio to UV interval plus
average fluxes and spectral indices in the X-ray and GeV range.

Clear systematic differences emerged between the three samples, in
particular concerning the slopes of the X-ray and gamma-ray emission.
Moreover the average luminosities of the three samples were different.
Since in each sample there are objects with different spectral properties
(e.g. X-ray to radio flux ratios) we thought that if luminosity was an
important parameter it would be useful to bin the objects not according
to their belonging to a specified sample, but according to
luminosity only.  The result is shown in Fig 1 (from Fossati et al. 1998). 
 Indeed, the resulting SEDs appear homogenous and
systematic trends are evident. For each luminosity class 
the derived SEDs show two very broad
components peaking between $10^{13} -10^{17}$Hz and between $10^{21}
-10^{24}$Hz respectively.  Analytic curves have been plotted for
comparison (see Fossati et al. 1998 for a complete account). The
underlying simple assumptions were: i) that the peak frequencies are
inversely related to the radio-luminosity.  ii) that the ratio of the two peak
frequencies in each SED is constant iii) that the height 
of the second peak is proportional to the radio luminosity.  
These "analytic" laws seem to represent the  average SEDs quite closely.
\begin{figure}
\hspace{1.3truecm}
\psfig{file=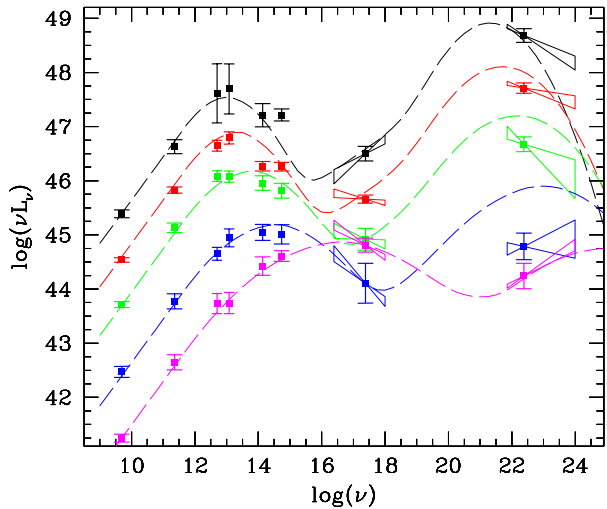,height=8.cm}
\vspace{-1.truecm}
\caption{Average SEDs of Blazars with different radio luminosity (from
Fossati et al. 1998).}
\end{figure} 

\subsection{Biases}

All the sources in the three samples have radio optical and X-ray data,
most have also infrared data, while only a fraction have gamma-ray data.
Within the three original samples, the fraction of objects detected in
gamma-rays are 19/50, 9/34 and 8/48 respectively.

Thus it is still an open issue whether the $\gamma$-ray properties of the
average SEDs are representative of the whole population or are
significantly biased. Clearly, since most sources are close to the
detection limit and many are known to be strongly variable
(e.g. Mukherjee et al. 1997), EGRET preferentially detected those that
were in an active state.  In an interesting but rarely quoted paper,
Impey (1996) showed via MonteCarlo simulations that, assuming all FSQ in
the 1 Jy sample have the same average $L_{\gamma}/L_{\rm radio}$ ratio,
the properties of the $gamma$-ray {\it detected} sample can be reproduced if
the intrinsic average $<L_{\gamma}/L_{\rm radio}>$ is about 1/10 of the
actually observed one, with a "normal" scatter of about 1 order of
amgnitude (the scatter could be due to variability).  We did not apply
any correction to the SEDs because the estimate though interesting is
rather uncertain. Moreover the analysis of Impey refers to FSQ and it
would not be justified to assume the same correction factor for all
luminosity classes.  There can be little doubt however that the average
gamma ray fluxes in Fig. 1 are overestimated.

In order to assess the possible differences between blazars detected and
non detected in gamma-rays we obtained BeppoSAX observations of "non
gamma-ray" HPQs. A good example is 1641+399 (3C 345): the source is
clearly detected up to the PDS range with a hard X-ray luminosity
comparable to that in the first peak (Tavecchio et al., in prep). 
Thus the hypothesis that all
blazars emit gamma-rays at an average level not much different than those
actually detected with EGRET (within an order of magnitude) finds
support.

\subsection{The Unified Framework for the SEDs of blazars.}

Fig 1. strongly suggests that the same radiation mechanisms operate in
all blazars. The SEDs define a spectral sequence ( we will call red and
blue the objects at the different extremes of the sequence), implying
that the jet properties change with continuity, therefore there are no
qualitative differences among jets of different power. Beamed synchrotron
and inverse Compton emission from a single population of electrons
account very well for the observed SEDs except in the radio to mm range
where effects of selfabsorption and inhomogeneity are important (see
also Kubo et al. 1998 ). This model predicts that variability of the two
spectral components should be correlated especially at frequencies 
near the peaks. Given the difficulty of getting adequate data,
it is remarkable that this has been verified at least in some 
well studied objects. In the following we will assume that 
this emission model holds in general.

There are however quantitative differences between jets producing the
SEDs depicted in Fig 1..  Adapting model spectra to the observed SEDs of
about 50 sources with gamma-ray information (but in many cases rather
poor broad band data) Ghisellini et al. 1998 derived the physical
parameters of the jets including seed photons of internal (SSC) as well
as external (EC) origin.  The model fits suggest that the observed trends
are due to i) an increasing importance of external seed photons with
increasing jet luminosity ii) a decreasing "critical" energy of the
radiating electrons with increasing (total) radiation energy density.
The latter dependence is physically plausible since the radiation energy
density determines the energy losses of relativistic particles.  If the
critical electron energies were determined by a balance between
injection/acceleration and cooling processes the latter dependence could
be understood.

This "unified" theoretical scheme, attractive as it is, needs to be
tested in many respects. One can think of at least two ways of doing so :
i) determine the physical parameters in individual objects with good data
ii) understand the mechanisms of particle acceleration and injection from
detailed variability studies.

More indirectly, if the shapes of blazar SEDs are "biunivocally" related
to luminosity (of course some scatter will have to be allowed for) there
must be predictable consequences on number counts and luminosity
functions of different types of objects.  Even if FSQs and BL Lacs
contain "similar" jets (at least close to the nucleus) as suggested by
the continuity of the SEDs, we still need to understand the differences in
emission line properties.  Also in this respect continuity could hold in
the sense that the accretion rate may decrease continuously along the
sequence but the emission properties of the disk may not simply scale
with the accretion rate.

\section{Studies of individual objects.}

The detailed study of the spectral and variability characteristics of
single Blazars is an important approach complementary to the study of
samples described above. In the following we will focus on observations
obtained with the Italian-Dutch satellite BeppoSAX.

Unfortunately in the case of red blazars the study of the synchrotron
component is difficult, because of the position of the Synchrotron peak,
which falls in the poorly covered IR - FIR range. Furthermore the study
of the gamma-ray component in the MeV-GeV region of the spectrum has been
difficult in the last few years due to the loss of efficiency of EGRET
and is now impossible after reentry of CGRO.

We show in Fig. 2 the case of 3C 279, observed with BeppoSAX in January
1997 simultaneously with CGRO (Hartman et al. 2000, ) and close in time
(December 1996) with ISO (Haas et al. 1998). The source was found to be in a
rather low state, analogous to the low state observed in 1993 (also
shown). Also shown for comparison are the two highest states recorded in
1991 and 1997.

A general correlation of the optical X-ray and $\gamma$-ray fluxes is
apparent.  The FIR data strongly suggest an additional, highly luminous
thermal dust component,
presumably observed at low inclination with respect to a putative
accretion disk. In any case the synchrotron peak is difficult to
localize.

\begin{figure} 
\plotone{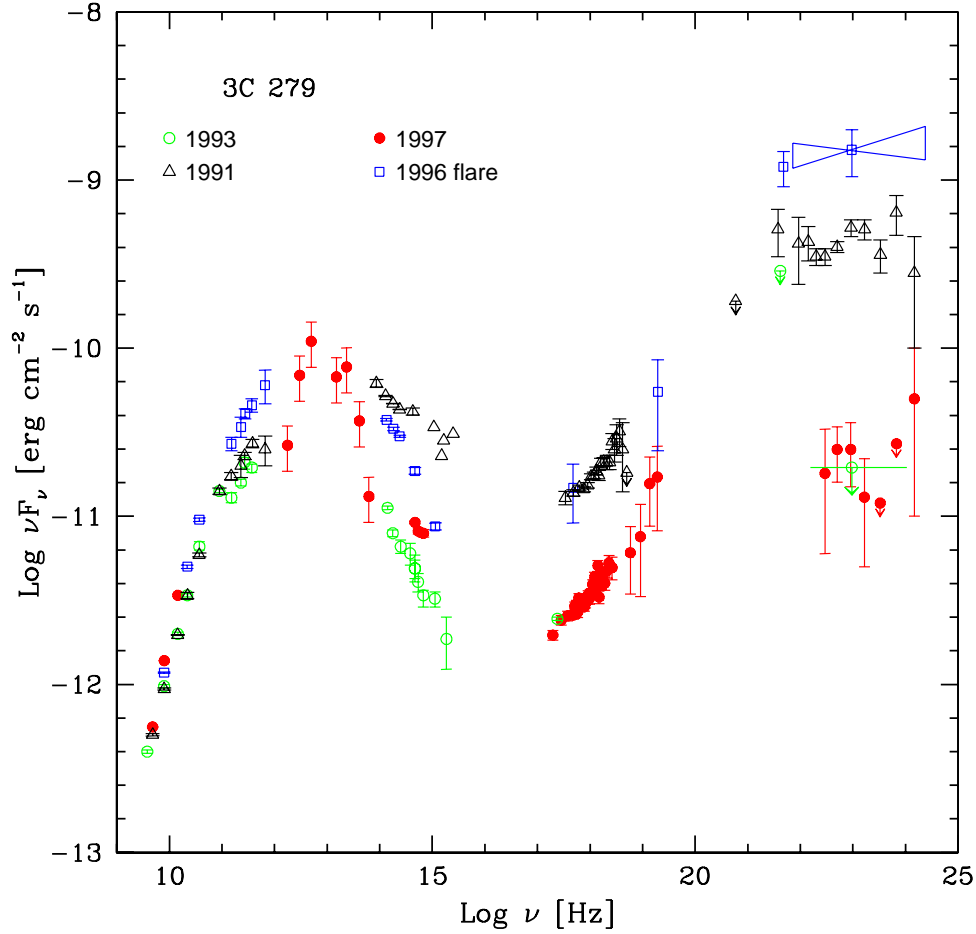}
\caption{Quasi-simultaneous SEDs of the quasar 3C279 taken in the
different epochs. The {\it Beppo}SAX and EGRET data taken in 1997 are
almost exactly contemporaneous, while the ISO spectrum is taken one month
before.}
\end{figure}

The situation is better for blue blazars. In several sources of this
class the Synchrotron component peaks in the X-ray band, where numerous
satellites can provide good data. In few bright extreme BLLac objects the
high energy $\gamma$-ray component is observable from ground with TeV
telescopes (for a general account see Catanese \& Weekes 1999). In these
particular cases the contemporaneous X-ray/TeV monotoring demonstrated
well the correlation between the Synchrotron and the IC components. A
very good example is Mkn 421 for which we obtained the observation of a
simultaneous TeV/X-ray flare with Whipple and {\it Beppo}SAX in 1998,
probing for the first time the existence of correlation on short (hour)
time scales (Maraschi et al. 1999; see also Takahashi et al. 1999,
Catanese \& Sambruna 2000). When the position of the two peaks can be well
determined observationally, as is possible in this type of sources,
robust estimates of the physical parameters of the jet can be obtained
(e.g. Tavecchio et al. 1998). This was done for both Mkn 421 and Mkn 501;
the case of Mkn 501 is shown in Fig 3a.

The broad band response of BeppoSAX allows a reliable determination of
the position of the synchrotron peak, $E_{\rm peak}$, if it falls between
0.1 and 100 keV.  We could verify that during the flare of Mkn 421
mentioned previously $E_{\rm peak}$, moved to higher energies with
increasing intensity (Fossati et al. 2000). The same behaviour was
exhibited in a more dramatic way by Mkn 501. Its synchrotron peak
moved to $E>100$ keV during the extraordinary activity in April
1997. Subsequent snapshot spectra obtained with BeppoSAX showed a
systematic decrease of $E_{\rm peak}$ down to $\simeq 0.1$ keV in June
1999 while the source was fading (Fig. 3b, Tavecchio et al. in
preparation). Over this two year period the X-ray light curve as measured
by the ASM aboard XTE was not monotonic with an overall decay interrupted
by flares. Thus our observations show that $E_{\rm peak}$ correlates with
luminosity not only along individual flares but also on much longer
timescales.

\begin{figure} 
\plottwo{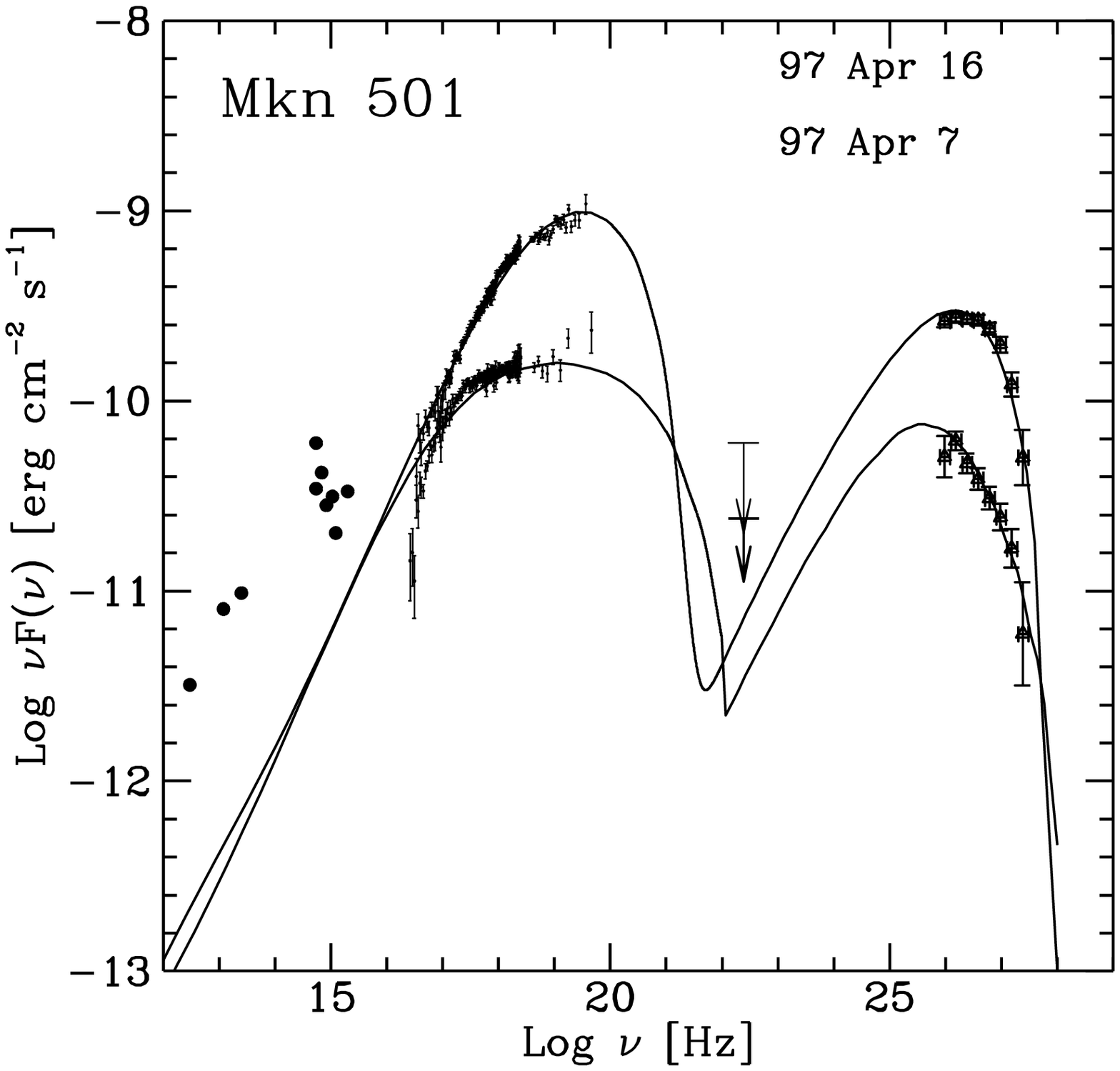}{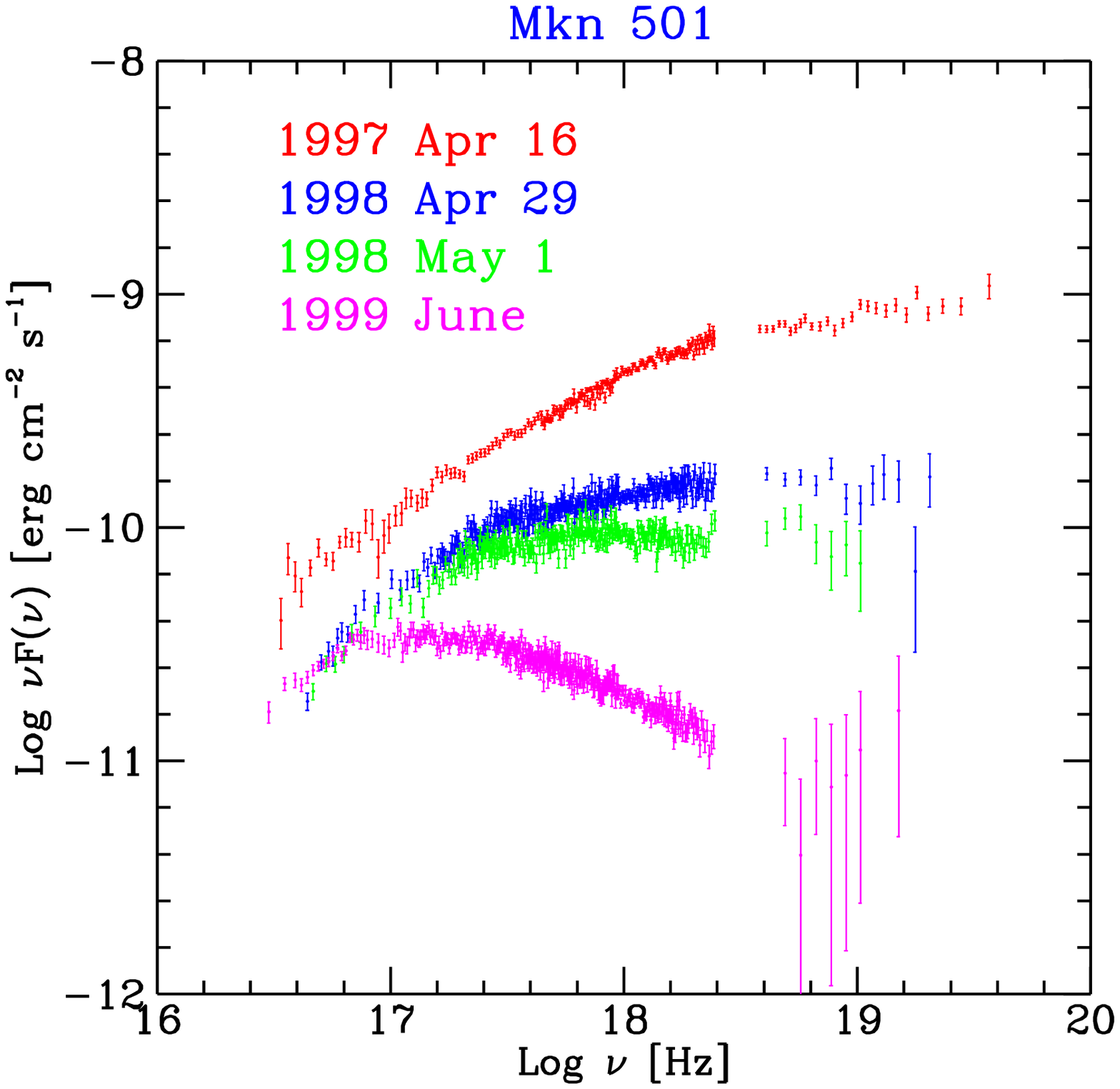}
\caption{{\it Left:} Overall SED of Mkn 501 observed simultaneously by
{\it Beppo}SAX and CAT during the majot flare in April 1997 (From
Tavecchio et al. 2000, in prep). The solid line is the spectrum obtained
with the SSC model. The possibility to constrain both the two peaks
allows to obtain robust estimates of the physical parameters of the
jet. {\it Right:} Spectral evolution of Mkn 501 during the period
1997-1999 (from Tavecchio et al. 2000, in prep). It is clear the strong
relation between the position of the peak and the total luminosity.}
\end{figure}

 The $E_{\rm peak}$ vs. luminosity relation observed 
in the time dependent behaviour of these two sources (higher $E_{\rm peak}$
for higher luminosity) is  opposite to that found in
the ``spectral sequence'', where  the peak falls at {\it
lower} frequencies for objects of higher luminosity.

A general scenario that could include both types of behaviour is the
following.  Let us suppose that the Lorentz factor of particles emitting
at the peak $\gamma _{\rm p}$ is determined by the equilibrium between
the cooling and acceleration processes, namely $t_{\rm acc}(\gamma _{\rm
p})=t_{\rm cool}(\gamma _{\rm p})$. Given that $t_{\rm cool}={\rm
const}/U\gamma $, where $U=U_{\rm rad}+U_B$ is the total energy density,
and using the general expression $t_{\rm acc}(\gamma )=\gamma t_{\rm
o,acc}$, found in the theory of diffusive shock acceleration (see
e.g. Kirk et al. 1998) one can write:
\begin{equation}
\gamma _{\rm p}=\left( \frac{\rm const}{Ut_{\rm o,acc}} \right)^{1/2}
\end{equation}
\noindent
This expression is consistent with the correlation found by Ghisellini et
al. (1998), $\gamma _{\rm p}\propto U^{-1/2}$ provided that the
acceleration timescale is, on average, similar in all sources. 
 Flares in single sources can then be interpreted as due to the
temporaneous decrease of $t_{\rm o,acc}$ due to changes in the
physical process of acceleration. This scenario seems to apply quite
well to Mkn 501 (Tavecchio et al. 2000, in prep), for which it is
possible to reproduce the observed variability with the only change of
$\gamma _{\rm p}$.

\section{Jet power vs. accretion power}

Finally we wish to discuss some recent results on luminous blazars with
emission lines. They are at the high-luminosity end of the sequence, with
the Synchrotron peak in the FIR region. In these sources the X-ray
emission is believed to be produced through the IC scattering between
soft photons external to the jet (produced and/or scattered by the Broad
Line Region) and electrons at the low energy end of their energy
distribution. Thus measuring the X-ray spectra and adapting a broad band
model to their SEDs yields reliable estimates of the total number of
relativistic particles involved, which is dominated by those at the
lowest energies. This is interesting in view of a determination of the
total energy flux along the jet (e.g. Celotti et al. 1997, Sikora et
al. 1997). The "kinetic" luminosity of the jet can be written as
\begin{equation}
L_{\rm j}=\pi R^2 \beta c \,U \Gamma ^2
\end{equation}
(e.g. Celotti et al. 1997) where $R$ is the jet radius, $\Gamma$ is the bulk
Lorentz factor and $U$ is the total energy density in the
jet, including radiation, magnetic field, relativistic particles and
finally protons. If one assumes that there is 1 (cold) proton per
relativistic electron the proton contribution is usually dominant.
   
In high luminosity blazars the UV bump is often directly observed and/or
can be estimated from the measurable emission lines, yielding important
information on the accretion process. Thus the relation between accretion
power and jet power can be explored.

For three sources of this type we performed a detailed analysis of the
{\it Beppo}SAX data and modeled the overall SED (Fig. 4, Tavecchio et
al. 2000). For other 6 sources also observed with {\it Beppo}SAX a
similar analysis is in progress and the results reported here are still
preliminary.

\begin{figure} 
\hspace{2.0truecm}
\psfig{file=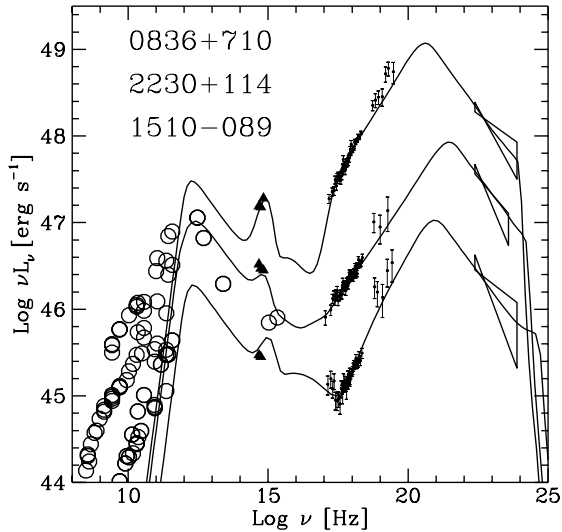,height=9.cm}
\vspace{-0.8truecm}
\caption{Overall SEDs of three powerful emission-lines Blazars (from
Tavecchio et al. 2000). The objects are characterized by the presence of
a strong UV-bumps, allowing the determination of the luminosity of the
accretion disk.}
\end{figure}

In all cases we estimated the kinetic luminosity of the jet and the
luminosity of the disk as described above. We can then compare the two in
Fig. 5b. We include also four BL Lac objects (namely BL Lac, ON231,
Mkn 501 and Mkn 421) for wich we have reliable information on the power
carried by the jet. Unfortunately we can set only upper limits on the
luminosity of their putative accretion  disks except for BL Lac, 
where the presence of a broad H$_{\alpha}$ line allows us 
to estimate of the ionizing continuum
(e.g. Corbett et al. 2000).

The diagram in Fig. 5a shows the comparison between the total radiative
luminosity $L_{\rm rad}$ of the jet and the power transported by the jet
including the proton contribution which is dominant. The ratio between
these two quantities gives directly the "radiative efficiency" of the
jet, which turns out to be $\eta\simeq 0.1$, though with large scatter.
The line traces the result of a least-squares fit: we found a the slope
$\sim 1$, indicating a rather constant radiative efficiency along the
Blazar sequence (note that the data cover a wide range of about 5 orders
of magnitude).

As discussed above we have reliable estimates of the luminosity released
by the accretion disk, calculated directly with the luminosity of the
blue bump or inferred by the luminosity of emission lines.  Therefore we
can directely compare the luminosity of the disk, $L_{\rm accr}$, and the
the luminosity of the jet, $L_{\rm jet}$. In Fig. 5b we report the
radiative luminosity of the jet $L_{\rm rad}$, which is a {\it lower
limit} of $L_{\rm jet}$ versus $L_{\rm accr}$. Note that in the case of
the four BL Lac objects the disk luminosity is formally an upper limit.
A similar approach was pioneered by Celotti et al. 1997, but their
estimates of $L_{\rm jet}$ were obtained applying the SSC theory to VLBI
data which refer to larger scales .

The first important result is that in some objects the power transported
by the jet is much {\it larger} (at least an order of magnitude) than the
luminosity released through accretion. This result is a strong challenge
for models elaborated to explain the formation of jets.

Two main lines of approach consider either extraction of rotational energy 
from the black hole itself or magnetohydrodynamic winds associated with 
the inner regions of accretion disks. We briefly discuss here the first case,
treated by  Blandford \& Znajek (1977). The result of their complex 
analysis is summarized in the well known expression:

\begin{equation}
L\simeq 10^{37}B^2 M_8^2 \,\,\,\,\, {\rm erg/s}
\end{equation}

Assuming maximal rotation for the black hole, the critical problem is the
estimate of the intensity reached by the magnetic field threading the
event horizon. Several authors have recently discussed this
difficult and subtle issue: here we use the results obtained by Ghosh \&
Abramovicz (1997) on the basis of equipartition within an accretion disk
described by the Shakura and Sunyaev (1973) model. Their estimates for
the rotational power are shown in Fig for various values of the mass of
the central black hole as a function of the luminosity observed from the
disk. The latter is related to the accretion rate which appears in the
formulae of Ghosh \& Abramovicz (1997) adopting an efficiency of
10\%. Clearly the model fails to explain the large power observed in the
jets of bright quasars, even for BH masses ($M\sim 10^9 M_{\odot}$).

In another class of models it is assumed that the energy powering the jet
is extracted directly from the accretion disk. Assuming a similar
efficiency, $\eta \sim 0.1$, for the conversion of both the accretion
power and the jet power into radiation, the plot in Fig. 5 suggests
that the mechanism responsible for the formation of the jet is able to
split the total accreted power $\dot M c^2$ in such a way that $L_{\rm
accr}\sim L_{\rm jet}$.

\begin{figure} 
\plottwo{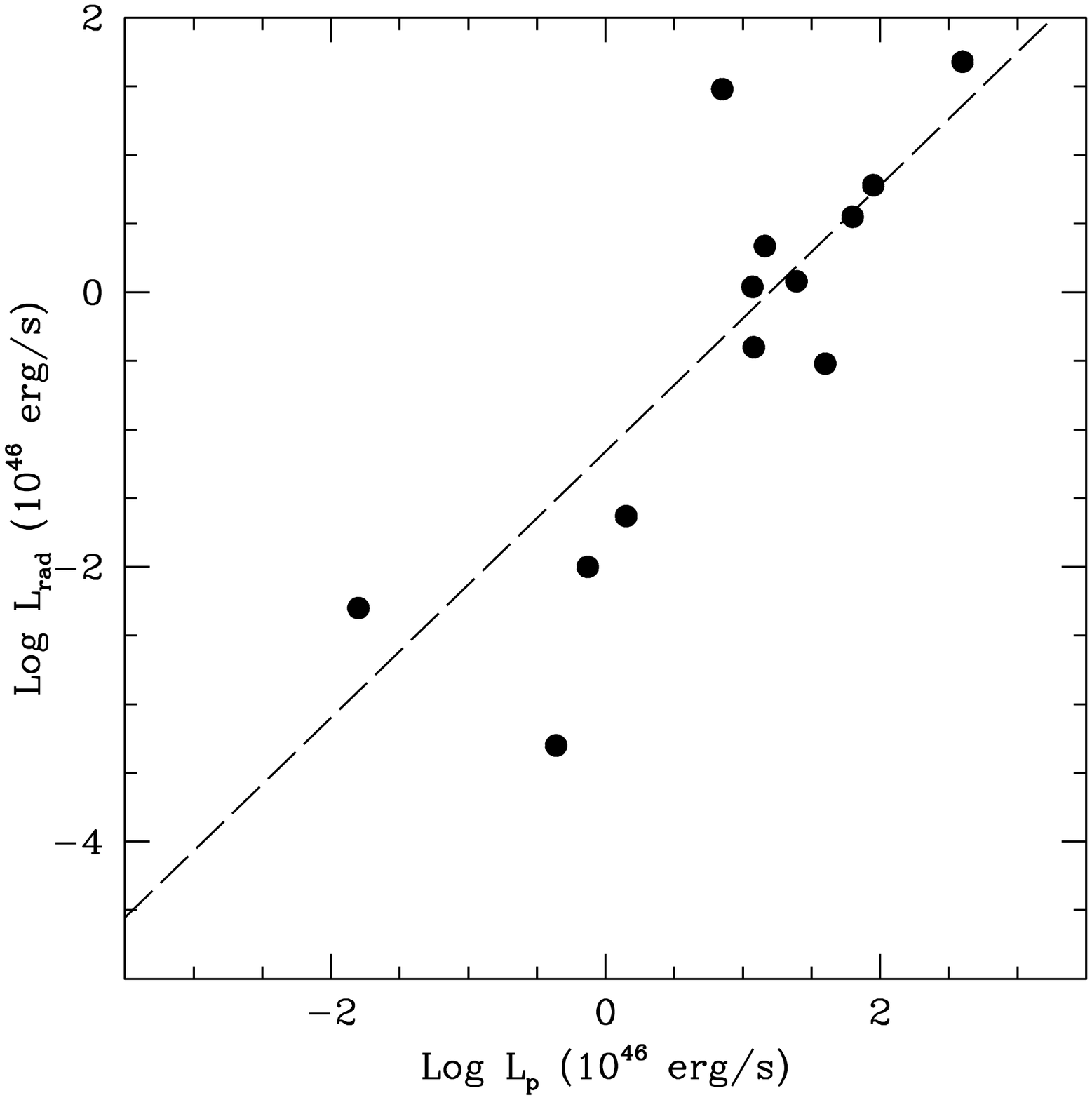}{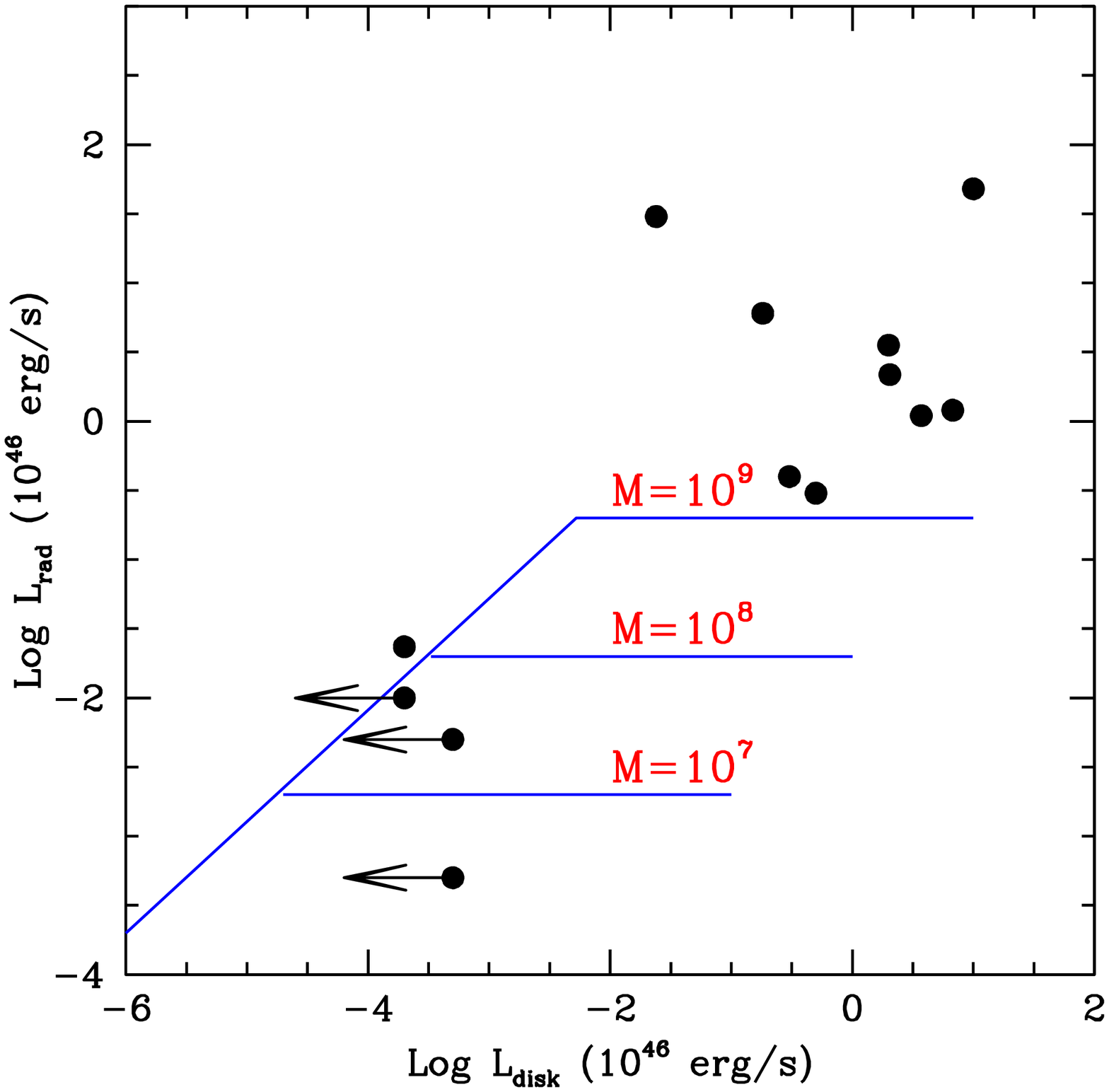}
\caption{{\it Left:} Radiative luminosity vs. jet power for the sample of
Blazars discussed in the text. The dashed line indicates the
least-squares fit to the data. {\it Right:} Radiative luminosity of jets
vs disk luminosity. The solid lines represent the {\it maximum} jet power
estimated for the Blandford \& Znajek model for black holes with
different masses (units are in Solar masses).}
\end{figure}

\section{Conclusions}

The study of broad band SEDs and their variability is essential for
understanding blazars. A unified approach is possible and valuable since
it can be tested and possibly disproved.  While the phenomenological
framework is suggested to be "simple" (e.g. "red" blazars are highly
luminous and emit GeV gamma-rays while "blue" blazars have low luminosity
and emit TeV gamma-rays) we do not yet know what determines the emission
properties of jets of different power nor what determines the jet power
in a given AGN.  There is however the exciting prospect that such
problems can be tackled with data that can be gathered in the near
future.

\section{References}

Blandford, R.D., \& Znajek, R.L., 1977, MNRAS, 179, 433\\
Blandford, R.D., \& Rees, M.J., 1978, Pittsburgh Conf. on BL Lac
Objects, p. 341-347.\\
Blandford, R.D. 1990, in Saas-Fee Advanced Course 20. Lecture Notes, 
Springer-Verlag\\   
Catanese, M., \& Weekes, T.C., 1999, PASP, 111, 1193\\
Catanese, M.\& Sambruna, R.\ M.\ 2000, ApJ, 534, L39\\ 
Celotti, A., Padovani, P., \& Ghisellini, G. 1997, MNRAS, 286, 415\\ 
Corbett, E. A., et al. 2000, MNRAS, 311, 485\\ 
Fossati, G., et al. 1998, MNRAS, 299, 433\\
Fossati, G., et al. 2000, ApJ, 541, 166\\ 
Ghisellini, G., et al. 1998, MNRAS, 301, 451\\
Ghosh, P., \& Abramowicz, M.A. 1997, MNRAS, 292, 887\\ 
Haas, M., et al. 1998, ApJ, 503, L109\\
Impey, C.,1996, AJ, 112, 2667\\
Kirk, J.G., Rieger, F.M., \& Mastichiadis, A., 1998, A\&A, 333, 452\\
Kubo, H., et al. 1998, ApJ, 504, 693\\
Landau, R., et al. 1986, ApJ, 308, 78\\
Maraschi, L., et al. 1999, ApJ, 526, L81 \\
Mukherjee, R., et al. 1997, ApJ, 490, 116\\ 
Sambruna, R.,M., Maraschi, L., \&Urry, C.M., 1996, ApJ, 463, 444\\
Shakura, N. I. \& Sunyaev, R. A. 1973, A\&A, 24, 337\\ 
Sikora, M., et al. 1997, ApJ, 484, 108\\  
Takahashi, T., et al. 1999, Astroparticle Physics, 11, 177\\ 
Tavecchio, F., Maraschi, L., \& Ghisellini, G., 1998, ApJ, 509, 608\\ 
Tavecchio, F., et al. 2000, ApJ, in press (astro-ph/0006443)\\

\end{document}